# Search Algorithms for Conceptual Graph Databases*

*Abdurashid Mamadolimov*
*Malaysian Institute of Microelectronic Systems, Malaysia*
rashid.mdolimov@mimos.my

**Abstract.** We consider a database composed of a set of conceptual graphs. Using conceptual graphs and graph homomorphism it is possible to build a basic query-answering mechanism based on semantic search. Graph homomorphism defines a partial order over conceptual graphs. Since graph homomorphism checking is an NP-Complete problem, the main requirement for database organizing and managing algorithms is to reduce the number of homomorphism checks. Searching is a basic operation for database manipulating problems. We consider the problem of searching for an element in a partially ordered set. The goal is to minimize the number of queries required to find a target element in the worst case. First we analyse conceptual graph database operations. Then we propose a new algorithm for a subclass of lattices. Finally, we suggest a parallel search algorithm for a general poset.

## 1 Introduction

Knowledge representation and reasoning has been recognized as a central issue in Artificial Intelligence. Knowledge can be represented symbolically in many ways. One of these ways, called conceptual graph (CG) representation, is graph formalism, that is, knowledge is represented by labelled graphs, and reasoning mechanism is based on graph homomorphism. All kinds of knowledge - ontology, facts, rules and constraints - can be represented by conceptual graphs.

In this paper, we consider a database composed of a set of conceptual graphs, representing some assertions about a modelled world. Using conceptual graphs and graph homomorphism a basic query-answering mechanism can be built based on semantic search. A query made to this base is itself a conceptual graph. An element $f$ of the database is a real answer candidate for query $q$ if and only if there is a homomorphism from $q$ to $f$. We note that graph homomorphism checking is an NP-Complete problem, that is, homomorphism check is an expensive operation. Therefore the main requirement for conceptual graph database organizing and managing algorithms is to reduce the number of homomorphism checks.

Graph homomorphism defines a partial order over conceptual graphs. A finite poset (of CGs) can be considered a database model for a conceptual graph database. Ordering, updating and retrieval are the main operations of organizing and managing of CG databases. All these operations can be represented by the more basic operations: *Searching* and *Finding Parents/Children*. In this paper we consider the searching operation only.

The problem of searching in partially ordered sets has recently received considerable attention. Motivating this research are practical problems in filesystem synchronization, software testing, information retrieval, and so on. In practical applications, the elements of partially ordered sets can be complicated and comparison of elements may be expensive. In the conceptual graph case, comparison of elements is equivalent to graph homomorphism check. Since graph homomorphism is an NP-Complete problem, CG comparison takes a lot of time and there is no hope to reduce this time. Therefore the efficiency of a search algorithm depends directly on the number of CG comparisons.

The binary search technique is a fundamental method for finding an element in a totally ordered set. As in the totally ordered case, the goal is to minimize the number of queries required to find a target element in the worst case. First we analyse conceptual graph database operations. Then we propose a new algorithm for a subclass of lattices. Finally, we suggest a parallel search algorithm for a general poset.

___________________________________________________________________
*This research is sponsored in part by the Science Fund Grant 06-03-04-SF0032 from MOSTI, Malaysia

## 2 Related Works

Conceptual graphs constitute formalism for knowledge representation. Conceptual graphs were introduced by Sowa [1] as a combination of existential graphs (Peirce, 1909) and semantic networks (Richens, 1956). The first book on conceptual graphs [2] applied them to a wide range of topics in artificial intelligence, computer science, and cognitive science. Since 1991, Conceptual Graphs have been mathematically developed by Chein and Mugnier [3]. One of the main requirements for knowledge representation formalism is to be logically founded, which should have two essential properties with respect to deduction in the target logic: soundness and completeness. For conceptual graphs the equivalent logic is First Order Logic (FOL). The FOL semantic was defined in [2], and the soundness of conceptual graph homomorphism with respect to logical deduction was shown. The first proof of the completeness result, based on the resolution method, is given by Chein and Mugnier [4].

The NP-Completeness of conceptual graph homomorphism checking was proven in [4]. Several algorithms have been proposed for checking conceptual graph homomorphism [5-9].

The simplest case of a partially ordered set is a totally ordered set. The binary search technique is an optimal algorithm for finding an element in a totally ordered set [10]. In [11], Mozes, Onak, and Weimann present a linear-time algorithm that finds the optimal strategy for searching a tree-like partially ordered set. Levinson [12] and Ellis [13] have proposed several modifications of breadth/depth first search technique using some simple properties of partially ordered sets. Dereniowski [14] proves that finding an optimal search strategy for general posets is hard and gives a polynomial-time approximation algorithm with sublogarithmic approximation ratio. In [15], a sorting problem in a partially ordered set is studied.

## 3 Preliminaries

### 3.1 Partially Ordered Set (Poset)

**Definition.** A *partial order* is a binary relation $"\leq"$ over a set $P$ which is reflexive, antisymmetric, and transitive, that is, for all $a, b$ and $c$ in $P$, we have that:

(i) $a \leq a$ (reflexivity);
(ii) if $a \leq b$ and $b \leq a$ then $a = b$ (antisymmetry);
(iii) if $a \leq b$ and $b \leq c$ then $a \leq c$ (transitivity).

A set $P$ with a partial order $"\leq"$ is called a *partially ordered set* or *poset* $(P, \leq)$. For any two elements $a$ and $b$ of a poset $(P, \leq)$ we use $a \geq b$ if $b \leq a$. Also we use $a < b$ ($a > b$) if $a \leq b$ ($a \geq b$) and $a \neq b$. Let $(P, \leq)$ be a poset. If $\top \in P$ such that $\forall x \in P, x \leq \top$, then $\top$ is called the *top* element of $P$ (the *bottom* element $\bot$ is dually defined); a poset does not necessarily have a top (bottom) element, but if it has a top (bottom) element then it is unique.

Let $(P, \leq)$ be a poset and $D$ be any non-empty, finite subset of $P$: $D \subseteq P$, $|D| = n$, where $n$ is a positive integer. We note that $(D, \leq)$ is also a poset. Let $x$ be an element of $P$: $x \in P$. The following definitions are useful for future description.

$Ancestors(x, D) = \{a \in D: a > x\}$;
$Descendants(x, D) = \{d \in D: d < x\}$;
$Parents(x, D) = \{p \in Ancestors(x, D): \forall a \in Ancestors(x, D), p \leq a\}$;
$Children(x, D) = \{c \in Descendants(x, D): \forall d \in Descendants(x, D), c \geq d\}$;
$Indeg(x, D) = |Parents(x, D)|$;
$Outdeg(x, D) = |Children(x, D)|$.

We use $Ancestor(x, D), Descendant(x, D), Parent(x, D)$, and $Child(x, D)$ for an element of $Ancestors(x, D), Descendants(x, D), Parents(x, D)$, and $Children(x, D)$ respectively.

### 3.2 Representation of Posets

For representation of a poset we use lists of descendants or lists of ancestors in special form. Let $c_1, \dots, c_n$ be elements of poset $D$. We define partial order over integers $D' = \{1, 2, \dots, n\}$ in the following way: for any $x, y \in D'$, $x \leq y$ if and only if $c_x \leq c_y$. Two lists are associated with an element $c_i \in D$, $i = 1..n$: *list of descendants* **D**$(i)$ and *list of ancestors* **A**$(i)$. The list of descendants includes $Outdeg(i, D')$ as a first component of the list and first $Outdeg(i, D')$ ancestors are $Children(i, D')$:

$$\mathbf{D}(i) = \big(Outdeg(i, D'); Child_1(i, D'), \ldots, Child_{Indeg(i,D')}(i, D'); Descendant_{Indeg(i,D')+1}(i, D'), \ldots \big).$$

The list of ancestors is dually defined.

### 3.3 Lattice

**Definition.** A *lattice* is a partially ordered set in which any two elements have a unique least upper bound (also called the *join*) and a unique greatest lower bound (also called the *meet*).

**Definition.** Let $(L, \leq)$ be a finite lattice. An element $j \in L$ is *join-irreducible* if $j$ has exactly one child.

## 4 Conceptual Graphs

### 4.1 Definitions

A *conceptual graph* is a labeled bipartite multigraph. One set of nodes is called the set $C$ of concept nodes, and the other set $R$ is called the set of relation nodes. If a concept $c$ is the $i$-th argument of a relation $r$ then there is an edge between $r$ and $c$ that is labeled $i$. Concept and relation nodes are labeled by two partially ordered sets. These pair of partially ordered sets is called *vocabulary*. An edge labeled $i$ between a relation $r$ and a concept $c$ is denoted $(r, i, c)$. Formal definition of a vocabulary and conceptual graph can be found in [3, see *Basic Conceptual Graph*].

**Definition.** Let $g$ and $h$ be two CGs defined over the same vocabulary with the node sets $C_g \cup R_g$ and $C_h \cup R_h$ respectively. A *homomorphism* $\pi$ from $g$ to $h$ is a mapping from $C_g$ to $C_h$ and from $R_g$ to $R_h$, which preserves edges and may decrease concept and relation labels, that is:

- $\forall (r, i, c) \in g, \ (\pi(r), i, \pi(c)) \in h,$
- $\forall e \in C_g \cup R_g, \ l_h(\pi(e)) \leq l_g(e).$

Let $g$ and $h$ be two CGs defined over the same vocabulary. We say $g \geq h$ if there is a homomorphism from $g$ to $h$. It is easy to show that the introduced order is transitive and reflexive, but it is not antisymmetric.

**Definition.** Two CGs $g$ and $h$ are said to be *hom-equivalent* if $g \geq h$ and $h \geq g$, also denoted $g \equiv h$. Let $G$ and $H$ be two hom-equivalence classes of CGs defined over the same vocabulary. We say $G \succcurlyeq H$ if $g \geq h$ for some $g \in G$ and $h \in H$. It is clear that it does not matter how $g$ and $h$ are chosen from $G$ and $H$, they can be any representatives of respective hom-equivalence classes. We note that the introduced order $\succcurlyeq$ is a partial order. Let $\Omega$ be the set of all hom-equivalence classes of CGs defined over a given vocabulary. Let $\Delta$ be a non-empty finite subset of $\Omega$. Without loss of generality, we can consider a set $D$ of representatives of elements of $\Delta$. We call it a *CG dataset*. We note that a CG dataset is also a partially ordered set. Let $Q$ be any element of $\Omega$, $Q \in \Omega$, and let $q$ be any representative of $Q$, $q \in Q$. We call $q$ a *query element*. A *query comparison* is a comparison between query element and an element of a dataset. Only query comparison needs to check graph homomorphism, but comparison of two elements of a dataset can be done without graph homomorphism.

### 4.2 Organizing and Managing Operations

There are three main operations: *Ordering*, *Updating* and *Retrieval*. We note that since a CG database is a knowledgebase, the inference operation can be also considered. However we limit ourselves to the first three operations. It is clear that *Ordering* (generating initial dataset) is consecutively *Inserting* of query CG into dataset. *Updating* can be done using *Inserting* and *Deleting* of query CG. *Retrieval* uses only *Finding Descendants* of query CG. Next representations look like:

- *Deleting* = *Searching* + something;
- *Inserting* = *Finding Parents* + *Finding Children* + something;
- *Finding Descendants* = *Finding Children* + something.

It can be easily checked that the "something" parts of the above representations do not depend on query comparisons and take insignificant time. *Finding Parents* and *Finding Children* are dual operations.

Therefore only two operations - *Searching* and *Finding Parents/Children* - are enough for manipulating a CG database. We consider only the *Searching* operation in a poset. The formal definition of the *Searching* problem can be seen in the input-output part of Algorithm 2.

## 5  Searching in Matryoshka-Lattices

There are different classes of posets. We can order poset classes according to how easy they are to search:
1. Totally ordered sets (chains);
2. Tree-like posets;
3. Lattices;
4. General posets.

In Figure 1 you can see examples of each class of posets. It is clear that for the searching operation the most amenable class is chains. Existing works show that the next best class is tree-like posets, then lattices, and the worst class is general posets. In the conceptual graph case we usually have lattices or general posets.

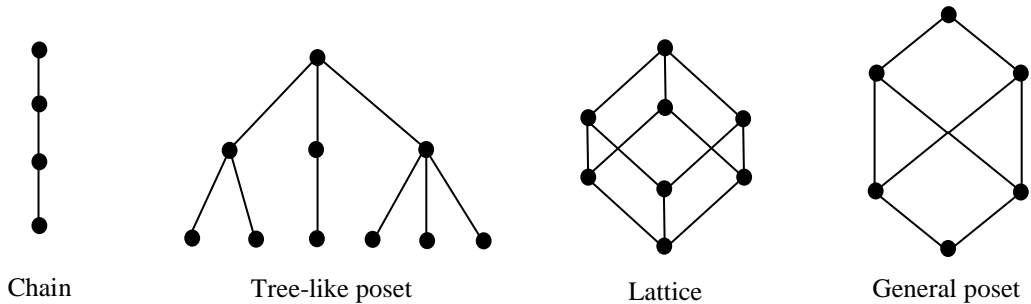

Chain    Tree-like poset    Lattice    General poset

Figure 1

Next we consider lattices. Let $(L, \leq)$ be a finite lattice. For each element $x \in L$ we define $J(x) = \{j \text{ join-irreducible}: j \leq x\}$.

**Theorem** [3]. Let $(L, \leq)$ be a finite lattice. For all $x, y \in L$, $x \leq y$ if and only if $J(x) \subseteq J(y)$.

Let $(L, \leq)$ be a finite lattice. We number join-irreducible elements of $(L, \leq)$ by 1,2,3, …,$|J(L)|$, where $J(L)$ is the set of join-irreducible elements of $L$. For each element $x$ of the lattice we can assign *binary code* of length $|J(L)|$: if the $i$-th join-irreducible element is less than or equal to $x$ then the $i$-th component of the binary code is 1, otherwise it is 0. Regarding the last theorem, it is clear that for all $x, y \in L$, $x \leq y$ if and only if the binary code of $x$ is less than or equal to the binary code of $y$. In Figure 3, lattice $L = L_0$ has six join-irreducible elements. We assigned a binary code of length 6 to each element of the lattice.

Habib and Nourine [16] have used binary codes for lattice operations. In the searching problem, if query element has already been compared with join-irreducible elements then it can be compared with other elements using binary code comparisons. For example, a Boolean lattice with $N = 2^n$ elements has $\log_2 N = n$ join-irreducible elements and this is enough query comparisons; all other comparisons can be done using binary code comparison.

Let $L$ be a finite lattice. Let $J(L)$ be the set of join-irreducible elements of $L$. We consider the set $L_1 = J(L) \cup \{\top, \bot\}$. The partially ordered set $L_1$ is called the *poset generated by join-irreducible elements of $L$*. We note that $L_1$ is not necessarily a lattice. In Figure 2 you can see a counterexample. If $L_1$ is a lattice then we can construct a poset $L_2$ generated by the join-irreducible elements of $L_1$. We can have a sequence $L = L_0, L_1, L_2, \ldots, L_{i-1}, L_i, \ldots$ of lattices, where $L_i$ is a poset generated by the join-irreducible elements of lattice $L_{i-1}$. We assumed that for all $i = 1,2,\ldots$, posets $L_i$ are lattices. If a lattice certifies the just explained property then we call it a *matryoshka-lattice*. The lattice $L = L_0$ in Figure 3 is a matryoshka-lattice, but the (left) lattice in Figure 2 is not. It is clear that $|L_{i-1}| \geq |L_i|$. There exists a positive integer $t$ such that all elements of the lattice $L_t$, except the bottom and may be the top elements, are join-irreducible. Let $t$ be a first such integer. Then any lattice $L_{t+p}$ is the same as lattice $L_t$. Now we have a finite sequence of mutually different lattices $L = L_0, L_1, L_2, \ldots, L_t$. It is clear that $L_t/\{\top\}$ is a tree-like poset and it has fewer elements than the original lattice $L$. It is enough to know comparisons between a query element and elements of $L_t/\{\top\}$. Therefore we have a more simple search space than in the original one. Using binary codes we can compare a query element with elements of $L_{t-1}$, then with elements of $L_{t-2}$, and so on, finally with elements of $L$. This is the idea of Algorithm 1.

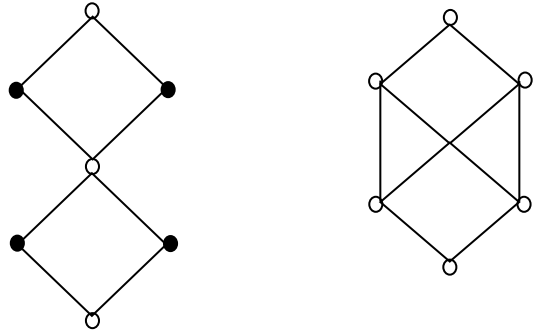

Figure 2

Algorithm 1 first constructs the sequence $L = L_0, L_1, L_2, \ldots, L_t$. Then using an efficient search algorithm for a tree, for example the algorithm from [11], a query element is compared with elements of $L_t/\{\top\}$. Using binary codes a query element is then compared with elements of the previous lattice, and so on. Finally Algorithm 1 finds comparisons between query element and elements of the original lattice $L$. Algorithm 1 is based on replacement of comparisons of two elements of the lattice with binary code comparisons. We note that comparison of binary codes is insignificant relative to comparison of CGs using graph homomorphism.

**Algorithm 1:** SearchInMat-Lattice(*Matryoshka-lattice, query element*);
**Input:**   Matryoshka-lattice $L$ and query element $q$;
**Output:** An element of matryoshka-lattice $L$, which is equal to query element $q$ or "No $q$ in a given matryoshka-lattice $L$";
1. $i = 0$;
2. $L_i = L$;
3. **If** $L_i$ has a non-join-irreducible element ($\neq \top, \bot$) **then**
    3.1 Find and numerate all join-irreducible elements of $L_i$;
    3.2 Assign binary code to each element of $L_i$;
    3.3 Construct next lattice $L_{i+1}$ generated by the join-irreducible elements of $L_i$;
    3.4 $i = i + 1$; **Go to** 3;
4. Using search algorithm in a tree ($L_i/\{\top\}$) find binary code of query element $q$;
5. **If** $i > 0$ **then** $i = i - 1$ else 7;
6. Extend binary code of query element $q$ using binary code comparisons in $L_i$; **Go to** 5;
7. Search for query element $q$ using binary code comparisons in $L$.

In Figure 3, lattice $L_1$ is generated by the join-irreducible elements of $L_0$, and lattice $L_2$ is generated by the join-irreducible elements of $L_1$. $L_2/\{\top\}$ is a tree-like poset. The bold nodes of $L_0$, $L_1$, and $L_2$ are join-irreducible elements. A binary code is assigned to each node of $L_0$ and $L_1$. For any given query element, Algorithm 1 uses only three query comparisons. These are comparisons between a query element and three join-irreducible elements of a tree-like poset $L_2/\{\top\}$.

## 6  Parallel Search Algorithm

Now we consider a general poset. First we propose a simple sequential search algorithm in a poset, and then using it we suggest a parallel algorithm.

### 6.1 Sequential Search Algorithm

Let $(P, \leq)$ be a poset and $D$ be any non-empty, finite subset of $P$: $D \subseteq P$, $|D| = n$, where $n$ is a positive integer. We call the set $D$ as a *dataset*. We assume that the dataset $D$ has both top and bottom elements and that they are different: $\top \neq \bot$. Assume $c_1, c_2, ..., c_n$ are elements of $D$ and $c_1 = \top$ and $c_n = \bot$. Let $q$ be an element of $P$: $q \in P$. We call the element $q$ a *query element*.

Our search algorithm is based on the following very simple property of posets: Let $x, y, z$ be elements of some poset. If $x \geq y$ is FALSE then for any $z$ such that $z \leq x$, we have that $z \geq y$ is also FALSE.

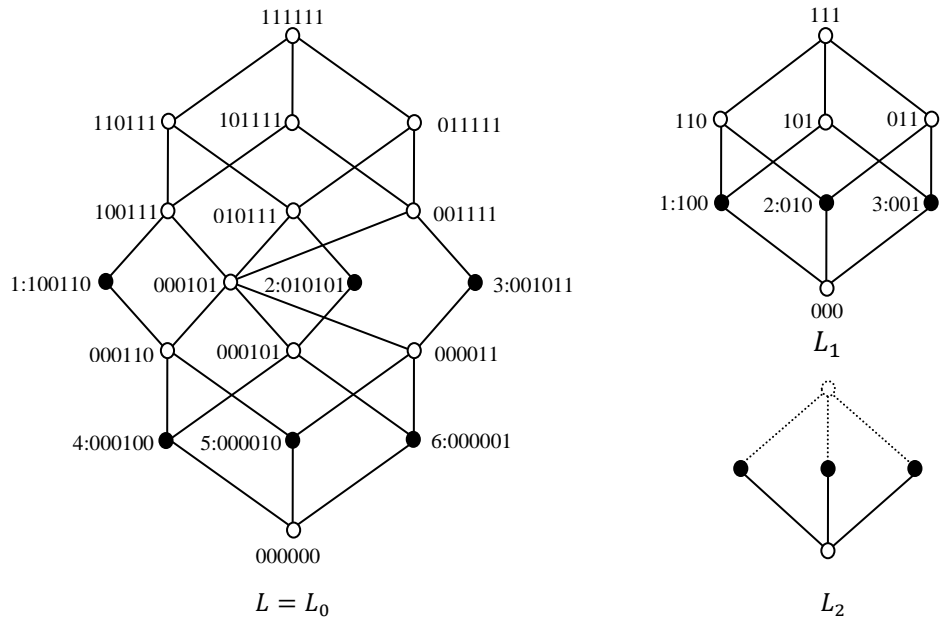

Figure 3

Algorithm 2 starts from the top element and moves to depth or width for further searching. Algorithm 2 asks a question of the form "*selected element* $\geq$ *query element*?". If the answer is "Yes" then it continues the search in the set of descendants of the selected element (moving to depth). If the answer is "No" then it searches in the complimentary part (moving to width).

**Algorithm 2:** SearchInPoset(*Dataset, query element*);
**Input:** Elements $c_i, i = 1..n$ of dataset $D$,
  Lists of descendants $\mathbf{D}(i)$ of elements $c_i$, $i = 1..n$ of dataset $D$,
  Query element $q$ ($\neq \bot$);
**Output:** If there exists an integer $x$ such that $1 \leq x \leq n$ and $q = c_x$ then $x$ else "No $q$ in $D$";

1. $x \leftarrow 1$; //*algorithm starts from top element*
2. Status($x$) $\leftarrow$ YES; //*if current element x is more or equal than query element then its status is YES*
3. $k \leftarrow 1$; //*starting from first child of current element*
4. $y \leftarrow Child_k(x, D')$; //*next child of current element is assigned to y*
5. **If** Status($y$) =NO **then go to** 7; //*status of child may already be NO*
6. **If** $c_y \geq q$ **then** $x \leftarrow y$ and **go to** 2 **else go to** 8; //*if child more or equal than query then child is current element and go ...*
7. **If** $k \neq Outdeg(x, D')$ **then** $k \leftarrow k + 1$ and **go to** 4 **else go to** 9; //*if y is not the last child of the current element then with next child go ...*
8. **If** $k \neq Outdeg(x, D')$ **then**
   Status($y$) $\leftarrow$ NO and
   **for all** $Descendant(y, D')$ **do** Status($Descendant(y, D')$)$\leftarrow$NO and
   $k \leftarrow k + 1$ and **go to** 4; // *because child is not more or equal than query (step 6) status of child and all its descendants is NO and with next child go ...*
9. **If** $c_x \leq q$ **then go to** 11; //*if current element is less or equal than query then go ...*
10. **Return** "No $q$ in $D$" and **stop**;
11. **Return** $x$;

We note that the $Child_k(x, D')$ function returns the $k$-th child of $x$ in $D'$. Because $x \in D'$, the function can be done in constant time since it corresponds to an access to an element of the descendant list $\mathbf{D}(x)$. The $Descendant(y, D')$ function is a similar. Also we note that the dual algorithm of Algorithm 2 can be easily obtained. It can be used if it is known that the query element is closer to the bottom element.

### 6.2 Parallel Search Algorithm

For parallel computing we use the parallel random-access machine (PRAM) model of computation with $P_1, P_2, \ldots, P_m$, $m \geq 2$ processors. In the PRAM model of computation all processors can access common a RAM in a single algorithm-step.

In parallel Algorithm 3, the first processor uses Algorithm 2; all other processors use Algorithm 2 with random chosen initial element (step 1.3). Shared information for processors is the Status of an element, which can be ANC (ancestor), NONANC (non-ancestor), and DES (descendant). In Algorithm 3 any query comparison (in steps 1.4, 1.10, and 1.14) will be done once only, that is, two or more processors do not check the same query comparison if they do not come to it at exactly the same time. In other words, there is no overlapping of partition of search space up to query comparisons. The whole algorithm stops when one of the processors $P_s$ returns $x_s$ or the first processor $P_1$ returns "No $q$ in $D$". If the processor $P_s$, $s \neq 1$ reaches the bottom element, but cannot find a searching element then it chooses a new initial element and searches again. This provides uniform distribution of tasks among the processors.

**Algorithm 3:** ParallelSearchInPoset(*Dataset, query element*);
**Input:** Elements $c_i, i = 1..n$ of dataset $D$,
Lists of descendants $\mathbf{D}(i)$ of elements $c_i$, $i = 1..n$ of dataset $D$,
Query element $q$ ($\neq \perp$);
**Output:** If there exists an integer $x_s$ such that $1 \leq x_s \leq n$ and $q = c_{x_s}$ then $x_s$
else "No $q$ in $D$";

1. **For all** processors $P_s$, $s = 1..m$ **do in parallel**
    1.1. **If** $s = 1$ **then** $x_s \leftarrow 1$ and **go to** 1.5;
    1.2. $R \leftarrow \{2, 3, \ldots, n - 1\}$;
    1.3. $x_s \leftarrow rand(R)$ and **if** Status($x_s$)=ANC **then**
        $R \leftarrow \{Descendants(x_s, D')\}$ and **go to** 1.3
        **elseif** Status($x_s$)=NONANC **then**
        $R \leftarrow R/(\{x_s\} \cup Descendants(x_s, D'))$ and **go to** 1.3;
    1.4. **If** $\neg(c_{x_s} \geq q)$ **then do**
        Status($x_s$)←NONANC and
        **for all** $Descendant(x_s, D')$ **do** Status($Descendant(x_s, D')$)←NONANC and
        $R \leftarrow R/(\{x_s\} \cup Descendants(x_s, D'))$ and **go to** 1.3;
    1.5. Status($x_s$)←ANC;
    1.6. $k_s \leftarrow 1$;
    1.7. $y_s \leftarrow Child_{k_s}(x_s, D')$;
    1.8. **If** $s \neq 1$ and Status($y_s$)=ANC **then** $R \leftarrow \{Descendants(y_s, D')\}$ and
        **go to** 1.3;
    1.9. **If** Status($y_s$) =NONANC **then go to** 1.11;
    1.10. **If** $c_{y_s} \geq q$ **then** $x_s \leftarrow y_s$ and **go to** 1.5 **else go to** 1.12;
    1.11. **If** $k_s \neq Outdeg(x_s, D')$ **then** $k_s \leftarrow k_s + 1$ and **go to** 1.7 **else go to** 1.13;
    1.12. **If** $k_s \neq Outdeg(x_s, D')$ **then**
        Status($y_s$) ← NONANC and
        **for all** $Descendant(y_s, D')$ **do** Status($Descendant(y_s, D')$)←NONANC and
        $k_s \leftarrow k_s + 1$ and **go to** 1.7;
    1.13. **If** Status($x_s$)=NONDES **then go to** 1.16;
    1.14. **If** $c_{x_s} \leq q$ **then return** $x_s$ and **go to** 2;
    1.15. Status($x_s$)←NONDES;
    1.16. **If** $s = 1$ **then return** "No $q$ in $D$" and **go to** 2 **else go to** 1.3;
2. **End.**

# 7 Conclusions

We analyzed conceptual graph database organizing and managing operations. We proposed a search algorithm for a particular case, when the search space is a special subclass of lattices, so called matryoshka-lattices. In the general case we suggested a parallel search algorithm which is based on simple sequential algorithm.

Since the searching in lattices problem is replaced with searching in a tree-like poset with fewer elements than original one we believe that Algorithm 1 is very efficient. However, we do not know how large/small the class of matryoshka-lattices is. There is no easy way to check whether a given lattice is a matryoshka-lattice or not.

We remark that the efficiency of Algorithm 3 can be increased by replacing the basic sequential algorithm (Algorithm 2) with a faster one.

# 8 Acknowledgements

I would like to thank Dickson Lukose for posing the problem. I thank K. Arichandran A/L M.S. Kandiah and Aziza Mamadolimova for discussions on these ideas. I also thank Chien Su Fong for managing the project. Finally, I would like to acknowledge Abdumalik Rakhimov for some suggestions.